\PassOptionsToPackage{expansion=false}{microtype}
\documentclass[sigconf]{acmart}

\usepackage[english]{babel}
\usepackage{booktabs}
\usepackage{multirow}
\usepackage{enumitem}
\usepackage{algorithm}
\usepackage{algpseudocode}

\setlist{nosep,leftmargin=*}
\AtBeginDocument{%
  \setlength{\abovedisplayskip}{4pt plus 1pt minus 1pt}%
  \setlength{\belowdisplayskip}{4pt plus 1pt minus 1pt}%
  \setlength{\abovedisplayshortskip}{3pt plus 1pt minus 1pt}%
  \setlength{\belowdisplayshortskip}{3pt plus 1pt minus 1pt}%
}

\copyrightyear{2026}
\acmYear{2026}
\setcopyright{cc}
\providecommand{\setcctype}[1]{}
\providecommand{\Description}[1]{}
\setcctype{by}
\acmConference[QuNet '26]{3rd ACM SIGCOMM Workshop on Quantum Networks and Distributed Quantum Computing}{August 17--21, 2026}{Denver, CO, USA}
\acmBooktitle{3rd ACM SIGCOMM Workshop on Quantum Networks and Distributed Quantum Computing (QuNet '26), August 17--21, 2026, Denver, CO, USA}
\acmDOI{10.1145/3833409.3833444}
\acmISBN{979-8-4007-2892-1/2026/08}


\begin{document}
\title[SATLOCK]{SATLOCK: Handover-Coupled Scheduling for
Weather-Resilient Quantum Key Distribution over LEO Constellations}

\author{Mohammad Arif Hossain}
\affiliation{\institution{Middle Tennessee State University} \city{Murfreesboro} \state{TN} \country{USA}}
\email{mohammadarif.hossain@mtsu.edu}

\author{Md Jafrin Hossain}
\affiliation{\institution{C2C Tech} \city{Mönchengladbach} \country{Germany}}
\email{hossain.mdjafrin@c2c-tech.com}

\renewcommand{\shortauthors}{Hossain and Hossain}

\begin{abstract}
\sloppy
Routing quantum keys over low Earth orbit (LEO) satellite constellations is harder than classical routing: satellite handovers couple consecutive scheduling decisions, stochastic cloud cover can silently zero a ground link, and finite-key effects can eliminate short, low-elevation passes entirely. We present SATLOCK, a handover-aware Quantum Key Distribution (QKD) routing framework that combines (i) a composite channel model incorporating atmospheric loss, pointing jitter, Markov cloud cover, decoy-state estimation, and finite-key correction; (ii) an integer linear program (ILP) giving a provable handover-aware throughput upper bound; and (iii) a decentralized deep Q-network (DQN) baseline for weather-adaptive online routing. We evaluate two contention regimes on a Walker constellation serving intercontinental demands. In low contention (16 satellites, 6 demands), the ILP delivers 1,311 Mbit while the strongest heuristics reach 95--96\% of ILP. In high contention (8 satellites, 12 demands), where handovers become binding, heuristics drop to 89.5\% of ILP. The DQN agent reaches 91.8\% and 84.6\% of ILP in the two regimes; it learns effective per-demand weather policies but is limited in aggregate by the lack of cross-demand coordination.
\end{abstract}

\maketitle


\section{Introduction}
\label{sec:intro}
\sloppy
Secure communication is under growing pressure from two directions: large-scale quantum computers threaten to break the public-key cryptography underpinning today's networks, while increasingly autonomous AI systems are being deployed into critical infrastructure with documented gaps between their safety requirements and actual behavior~\cite{MAH_containment_gap}, expanding both the attack surface and the value of compromised keys. Quantum key distribution (QKD) offers a response whose security rests on physics rather than computational assumptions, but delivering it at a global scale requires satellites. Routing cryptographic keys over low Earth orbit (LEO) satellite constellations is qualitatively harder than routing classical traffic. Quantum signals cannot be amplified or copied~\cite{wootters1982noclone}, so every lost photon is permanent. Each satellite pass lasts only a few minutes, stochastic cloud cover can silently zero a ground link mid-session, and finite-key overhead renders short, low-elevation passes nearly useless. Most critically, satellite \emph{handovers} reassign a satellite to a different demand between consecutive time slots. This incurs a pointing-and-acquisition penalty that couples adjacent scheduling decisions in a way classical networks do not face~\cite{mehic2020quantum}. Ignoring this temporal coupling, as existing heuristics do, results in measurable throughput loss that increases with constellation contention.

QKD exploits quantum no-cloning to generate information-theoretically secure shared keys~\cite{bennett1984bb84}. LEO satellite links traverse the thin upper atmosphere, enabling intercontinental key exchange beyond the reach of terrestrial fiber, as demonstrated by the Micius mission~\cite{liao2017satellite}, and are now considered the most credible near-term path to a global quantum network~\cite{pirandola2020advances}. These satellite links are also expected to become a core layer of next-generation (6G and beyond) space-air-ground integrated networks (SAGIN), where quantum and classical traffic increasingly share the same orbital and ground infrastructure. As with other resource-allocation problems in such large, dynamic, heterogeneous networks, machine learning and deep reinforcement learning (DRL) have emerged as practical complements to exact optimization~\cite{MAH_FEEL, Liu_DT}, offering fast, adaptive decisions in settings where re-solving a combinatorial program at every time slot is infeasible. Yet no prior routing work simultaneously accounts for finite-key effects, stochastic weather, and handover-coupled scheduling. These three effects recur throughout the paper and merit precise definitions. \emph{Finite-key} effects are the security penalty incurred because a real pass yields only a finite number of detected signals, so the usable key rate falls below the asymptotic value and vanishes for very short passes~\cite{Dupuis_privacy}. \emph{Cloud cover} acts as an on/off mask: a clouded ground station receives no photons, regardless of geometry. A \emph{handover} is the reassignment of a satellite from one ground demand to another between consecutive slots, forcing the optical terminal to re-point and re-acquire and thus lose part of a slot, an effect that couples scheduling decisions across time.

We present \textsc{SATLOCK}, a handover-coupled scheduling framework for LEO QKD networks. Its core insight is that handover penalties create \emph{temporal coupling} between consecutive slot decisions: a globally optimal schedule avoids unnecessary satellite reassignments, while myopic per-slot heuristics do not. We quantify this gap with a provable ILP upper bound and show it reaches 10.5\% under high contention, a difference that matters for a technology where every secure bit is expensive. Complementing the ILP, \textsc{SATLOCK} includes a weather-adaptive DRL routing agent that makes per-slot decisions in microseconds, trading a modest optimality gap for real-time deployability. Together, the ILP benchmark and the learned policy span the accuracy--latency spectrum, providing both a certifiable planning tool for network operators and a lightweight controller suitable for onboard deployment. The remainder of this paper details the system and channel model, formulates the handover-coupled ILP, describes the DRL agent, and evaluates both against scenarios with varying contention and weather conditions.

\noindent\textbf{Contributions:}
\sloppy
\begin{enumerate}[leftmargin=*,topsep=2pt,itemsep=1pt]
  \item \textbf{Physically grounded channel model.} A composite link model combining Beer--Lambert absorption, diffraction, pointing jitter, Markov cloud cover, decoy-state estimation, and finite-key correction with satellite handover overhead, validated against a Qiskit BB84 simulation.
  \item \textbf{Handover-aware ILP upper bound.} An exact binary ILP, solved with CBC, that explicitly models the temporal coupling induced by satellite reassignment across slots, providing a tight optimality certificate unavailable in prior work.
  \item \textbf{Weather-adaptive online routing agent.} A per-demand DQN with progress-aware state and handover-penalized rewards, trained under Markov-correlated weather. At inference it decides in ${\sim}50\,\mu$s per slot without re-solving an NP-hard program, and its gap to the ILP under high contention quantifies the cost of independent per-demand coordination.
\end{enumerate}

\section{Related Work}
\label{sec:related}
\sloppy
Liao~et~al.~\cite{liao2017satellite} and Yin~et~al.~\cite{yin2020entanglement} demonstrated satellite QKD and entanglement distribution at intercontinental range, establishing the physical link model we adopt. Bedington~et~al.~\cite{bedington2017progress} and Ribezzo~et~al.~\cite{ribezzo2023deploying} identify pointing loss, cloud outages, and finite-key overhead as dominant impairments, with cloud-driven outages confirmed as the primary bottleneck in a deployed two-node network, but neither addresses multi-demand routing or handover scheduling. Mehic~et~al.~\cite{mehic2020quantum}, Cao~et~al.~\cite{cao2022evolution}, and Aguado~et~al.~\cite{aguado2019engineering} study trusted-node routing and SDN control for QKD over fiber, assuming deterministic links and asymptotic key rates. Caleffi~\cite{caleffi2017optimal} and Pant~et~al.~\cite{pant2019routing} propose fidelity- and success-probability-based routing for quantum repeater networks, assuming static link qualities without visibility windows or handover penalties. We reuse their criteria as baselines in Section~\ref{sec:eval}. Both stay within a few percent of optimal under low contention, but because they score each slot independently, they lose ground once handover coupling makes the best per-slot choice differ from the best schedule---precisely the regime this paper targets. \textsc{SATLOCK} is the first framework to jointly model finite-key correction, Markov cloud cover, and handover-coupled scheduling with a provable optimality certificate.

Classical LEO scheduling optimizes throughput under deterministic or i.i.d.\ channel models and ignores quantum-specific handover irreversibility: a handover penalty in a QKD slot cannot be recovered as classical retransmission allows. \textsc{SATLOCK}'s ILP captures this via temporal coupling constraints absent in classical scheduling. DRL has been applied to traffic engineering and wireless resource management~\cite{MAH_MEC_MRL, huang2022drl}, but without handover coupling or correlated quantum channels. Our DQN agent instead trains under Markov-correlated weather, includes a handover flag to expose coupling cost, and uses cumulative key progress for demand-aware reallocation, leaving the residual gap to the ILP as motivation for centralized multi-agent DRL.

\begin{figure*}[t]
  \centering
  \includegraphics[width=0.9\textwidth]{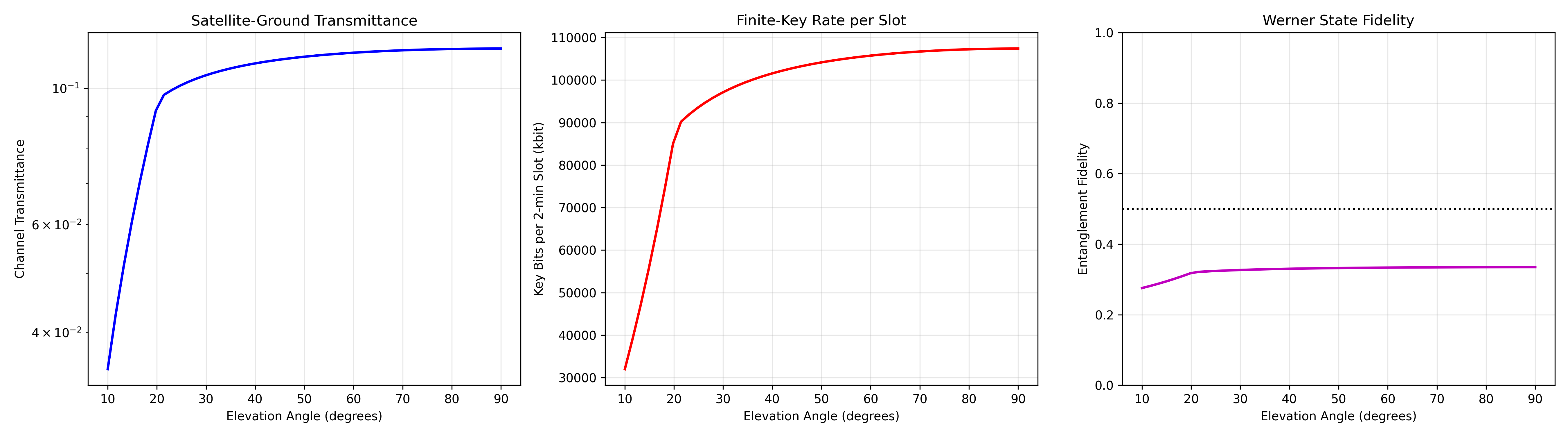}
  \Description{Schematic of the SATLOCK composite channel model combining atmospheric loss, pointing jitter, Markov cloud cover, decoy-state estimation, and finite-key correction.}
  \caption{Channel model effects. \emph{Left}: pointing-jitter loss.
    \emph{Center}: finite-key vs.\ asymptotic rate vs.\ pass duration.
    \emph{Right}: Werner-state entanglement fidelity vs.\ transmittance.}
  \label{fig:channel}
\end{figure*}

\subsection{Satellite Constellation and Ground Network}
We model a Walker 53°:16/4/1 constellation (inclination:total/planes/phasing notation): 4 orbital planes at 53° inclination, 500\,km altitude, 4 satellites per plane, $N_s=16$ total, with 90° inter-plane RAAN spacing. The ground network covers 10~stations across East Asia (Beijing, Tokyo, Seoul, Shanghai), South Asia (Mumbai), Southeast Asia (Singapore, Bangkok), and Europe (Vienna, London, Madrid). A satellite is \emph{visible} above 10° elevation ($\approx$2100\,km footprint radius); time is discretized into $T=60$ slots of 2~min over a 2-hour orbital window. This scale is deliberately modest: it keeps the ILP exactly solvable so that the optimality certificate is genuine, while still producing enough satellite--demand contention to expose handover coupling.

The LEO satellite layer serves as a quantum relay tier, forwarding key material between trusted ground nodes via direct or 1-hop relay paths. Each satellite is a \emph{trusted-node} relay rather than a quantum repeater: it terminates a downlink, measures the received key material, and re-transmits to the next hop, so no end-to-end entanglement is preserved across the relay. Each satellite also operates a single steerable downlink and therefore serves at most one ground station per slot. This is the physical origin of the exclusivity constraint in Section~\ref{sec:ilp}, where multi-beam payloads that lift this restriction are discussed as a relaxation.

\subsection{Composite Channel Model}
\label{sec:channel}
The end-to-end transmittance on a satellite-to-ground link combines four multiplicative terms:
\begin{equation}
  \eta_{\rm total} = \eta_{\rm atm}\;\eta_{\rm diff}\;\eta_{\rm jitter}\;\eta_{\rm weather},
  \label{eq:eta}
\end{equation}
where $\eta_{\rm atm}=e^{-\sigma_{\rm ext}\,z(\theta)}$ is Beer--Lambert extinction at slant angle $\theta$; $\eta_{\rm diff}=(D_r/(2\lambda L/D_t))^2$ is far-field diffraction loss; $\eta_{\rm jitter}=\exp(-2(\sigma_p/w_b)^2)$ is Gaussian beam-wander loss ($\sigma_p=0.7\,\mu$rad, Micius-class~\cite{liao2017satellite}, $w_b$ the beam waist radius); and $\eta_{\rm weather}\in\{0,1\}$ is a Markov-chain cloud mask with persistence $\rho=0.85$ and per-station clear-sky probability $1-p_c$ ($w_t=0$ blocked, $w_t=1$ clear). The transition matrix is:
\begin{equation}
  P(w_{t+1}\!\mid w_t) =
  \begin{pmatrix}
    1-p_c(1-\rho) & p_c(1-\rho) \\
    (1-p_c)(1-\rho) & 1-(1-p_c)(1-\rho)
  \end{pmatrix},
  \label{eq:weather}
\end{equation}
with stationary distribution $P(w=1)=1-p_c$. The \emph{finite-key} secret-key rate per slot is~\cite{tomamichel2012tight}:
\begin{equation}
  R_{\rm fk} = \begin{cases}
    \max\!\left(0,\; R_{\infty}
      - \dfrac{7\sqrt{\log(2/\varepsilon_s)}}{\sqrt{n_{\rm sig}}}\right)
      & \text{if } n_{\rm sig}\ge 10^4,\\[6pt]
    0 & \text{otherwise,}
  \end{cases}
  \label{eq:fk}
\end{equation}
where the prefactor~7 follows from the composable security analysis of~\cite{tomamichel2012tight} under collective attacks with $\varepsilon_s=10^{-10}$; $R_\infty$ is the asymptotic BB84 decoy-state rate~\cite{lo2005decoy}; and $n_{\rm sig}=f_{\rm rep}\,\eta_{\rm total}\,\eta_{\rm det}\,f_{\rm sift}\,\Delta t$ is the mean signal-photon count per slot. We use this form as a conservative, reproducible lower bound: it applies to collective attacks and inherits the asymptotic-equipartition-property (AEP) penalty, omitting both the finite sampling error in $R_\infty$ and the post-selection term needed for general (coherent) attacks. Tighter entropic-uncertainty bounds avoid the AEP loss and cover general attacks directly, raising the per-pass key volume; since every baseline in this paper consumes the same rate function, a tighter bound would shift all methods' throughput together without changing the relative comparison or the handover-coupling conclusion. A full entropic-uncertainty treatment is left to future work. Table~\ref{tab:gs} lists ground-station locations and $p_c$ values.

\begin{table}[t]
\caption{Ground station coordinates and cloud probabilities $p_c$
(MODIS Terra C6.1; $\rho=0.85$~\cite{pirandola2020advances}).}
\label{tab:gs}
\centering\scriptsize
\setlength{\tabcolsep}{6pt}
\begin{tabular}{lrrc|lrrc}
\toprule
\textbf{City} & \textbf{Lat} & \textbf{Lon} & \textbf{$p_c$} &
\textbf{City} & \textbf{Lat} & \textbf{Lon} & \textbf{$p_c$} \\
\midrule
Beijing  & 39.9N & 116.4E & 0.30 & Vienna & 48.2N & 16.4E & 0.45 \\
Shanghai & 31.2N & 121.5E & 0.45 & London & 51.5N &  0.1W & 0.60 \\
Tokyo    & 35.7N & 139.7E & 0.40 & Madrid & 40.4N &  3.7W & 0.30 \\
Seoul    & 37.6N & 127.0E & 0.35 & Mumbai & 19.1N & 72.9E & 0.65 \\
Singapore&  1.3N & 103.8E & 0.70 & Bangkok& 13.8N & 100.5E& 0.60 \\
\bottomrule
\end{tabular}
\end{table}

Figure~\ref{fig:channel} illustrates the three channel effects. Pointing jitter $\sigma_p=0.7\,\mu$rad is modeled as elevation-independent, following the Micius system characterization in~\cite{liao2017satellite}, where beam waist and slant range scale identically with the elevation angle, leaving $\eta_{\rm jitter}$ constant; elevation-dependent jitter modeling is deferred to future work. The finite-key correction becomes small relative to $R_\infty$ for passes exceeding 120\,s, but eliminates sub-60\,s passes entirely.

\subsection{Qiskit BB84 Validation}
We validate the channel model using a prepare-and-measure BB84 simulation in Qiskit with depolarizing noise~\cite{qiskit2024}. For each channel error rate in $[0, 0.20]$, we simulate 500 qubit transmissions per trial over 5 independent trials. The measured quantum bit error rate (QBER) tracks the theoretical relationship $\text{QBER}\approx e_{\text{channel}}$ within the binomial sampling uncertainty of the finite-shot simulation: at $\text{QBER}=5\%$, the per-trial standard error is $\sqrt{0.05\times0.95/500}\approx 1.0\%$, falling to $\approx0.44\%$ when averaged over five trials. We treat the Qiskit run as a protocol-level consistency check rather than as a model of QKD hardware, since a noisy intermediate-scale quantum (NISQ) device and a real photonic transmitter differ substantially; the simulation only confirms that the prepare-and-measure protocol reproduces the expected QBER-versus-error relationship assumed by our channel model. The finite-key rate~(\ref{eq:fk}) computed from simulated QBER remains within 1--2\% of the analytical prediction for passes above 60\,s, confirming end-to-end consistency between the protocol layer and the network-level channel model. All subsequent network-level results, therefore, rest on a rate function whose protocol-layer behavior has been independently reproduced rather than assumed.

\section{Problem Formulation and ILP Bound}
\label{sec:ilp}
Let $\mathcal{K}$ be the demand set and $\mathcal{P}_k$ the feasible paths for demand $k$ (direct and 1-hop relay). Binary variable $x_{k,t,p}\in\{0,1\}$ indicates whether demand $k$ uses path $p$ at slot $t$. To model pointing and acquisition overhead, a satellite incurs a handover loss when it serves different demands in consecutive slots; we use $y_{k,s,t}\in\{0,1\}$ to indicate that demand $k$ uses satellite $s$ in slot $t$, and $h_{s,t}\in[0,1]$ to indicate a handover on satellite $s$ between slots $t-1$ and $t$.

\paragraph{Objective.} Maximize delivered secret-key bits after handover loss:
\begin{equation}
  \max_{\mathbf{x},\mathbf{y},\mathbf{h}}\sum_{k,t,p}
    x_{k,t,p} R_{k,t,p}
    - \alpha \sum_{s,t} h_{s,t} R^{\max}_{s,t},
  \label{eq:obj}
\end{equation}
where $R_{k,t,p}$ is the finite-key secret-key volume on path $p$ at slot $t$, $R^{\max}_{s,t}$ is the maximum key volume involving satellite $s$ at slot $t$, and $\alpha=0.5$ models a 50\% throughput loss during a handover slot.

\paragraph{Constraints.}
Five constraint families encode the physical and logical structure of the problem.\\
\textbf{C1}: Each demand transmits over at most one path per slot, $\sum_p x_{k,t,p}\le 1\;\forall k,t$, so a demand cannot split its key traffic across multiple simultaneous routes.\\
\textbf{C2}: Each satellite serves at most one demand per slot, $\sum_{k,p\ni s}x_{k,t,p}\le 1\;\forall s,t$. This half-duplex exclusivity constraint reflects the operational mode of current Micius-class transmitters, which require dedicated optical alignment to a single ground station per time slot~\cite{liao2017satellite}; future multi-beam payloads would relax C2 to a per-satellite capacity bound $\sum_{k,p\ni s}R_{k,t,p}\,x_{k,t,p}\le C_{\max}$.\\
\textbf{C3}: A handover is detected whenever a satellite serves a demand in slot $t$ after serving a different demand in slot $t-1$, enforced by $h_{s,t}\ge y_{k,s,t}+\sum_{k'\ne k}y_{k',s,t-1}-1\;\forall k,s,t>1$, which forces $h_{s,t}$ to 1 exactly when both conditions hold.\\
\textbf{C4}: The path and satellite-usage variables are binary, and the handover indicator is bounded, $x_{k,t,p}\in\{0,1\}$, $y_{k,s,t}\in\{0,1\}$, $0\le h_{s,t}\le 1$; since handovers are penalized in (\ref{eq:obj}), $h_{s,t}$ takes binary values at optimality without an explicit integrality requirement.\\
\textbf{C5}: The satellite-usage indicators are tied to the path selection by $y_{k,s,t}=\sum_{p\ni s}x_{k,t,p}\;\forall k,s,t$, so $y_{k,s,t}=1$ exactly when the path chosen for demand $k$ at slot $t$ traverses satellite $s$.

\paragraph{Complexity.} The ILP has $O(K \cdot T \cdot \bar{P}+KST)$ binary variables, where $\bar{P}$ is the average number of feasible paths per demand-slot pair, plus $O(ST)$ continuous handover indicators, with constraint count $O(ST+KT+KST)$. The aggregate $y_{k,s,t}$ indicators avoid enumerating all cross-products of paths in adjacent slots, essential to keep the handover-aware formulation tractable. However, scaling to 100+ satellites would increase the variable count quadratically due to the relay path enumeration, motivating online learning policies that run at inference in ${\sim}50\,\mu$s per slot without re-solving. For our 16-satellite, 6-demand, 60-slot instance, the handover-aware ILP is solved in ${\sim}0.8$\,s by CBC~\cite{forrest2005cbc}.

\section{Learning-Based Online Routing Agent}
\label{sec:drl}
\subsection{Environment}
We implement \texttt{SatQKDEnv}, a gym-style environment with Markov weather, in which the agent observes at each slot:
\begin{equation}
  \mathbf{s}_t = \bigl[t/T,\;\mathbf{V}_t,\;\mathbf{w}_t,\;\mathbf{b}_t,\;g_t\bigr],
\end{equation}
where $\mathbf{V}_t\in\mathbb{R}^3$ contains the best available finite-key rates for each path type (direct, 1-hop relay, no-path) adjusted for current weather; $\mathbf{w}_t\in\{0,1\}^2$ is the source/destination weather state; $\mathbf{b}_t\in\mathbb{R}$ is the cumulative key volume delivered for the demand; and $g_t$ flags whether the currently available paths would reuse a satellite that served a different demand in the previous slot, giving a per-agent state dimension of $1+3+2+1+1=8$. The compact state is deliberate: normalized time, weather-adjusted rates, own progress, and a single handover flag suffice for the agent to trade off immediate key delivery against reassignment cost, without observing other demands directly.

\subsection{Action Space and Reward}
Each per-demand agent independently selects one of 3 routing options (skip / direct / 1-hop relay), yielding a combined action space of $3^K$ across all $K$ agents. Because agents observe only their own state and act without explicit coordination, conflicting satellite requests are possible; this decentralized structure is the source of the residual gap to the centralized ILP examined in Section~\ref{sec:eval}. Per-slot reward:
\begin{equation}
  r_t = \sum_k R_{k,t,\pi_k(a)} - \lambda_h H(a) - \lambda_g G(a),
  \label{eq:reward}
\end{equation}
where $H(a)=\sum_k h_k(a)$ is the total relay-hop count ($h_k(a)\in\{0,1\}$, the relay hops used by demand~$k$ under action~$a$) and $G(a)$ counts handovers caused by the selected actions; we set $\lambda_h=0.001$ and $\lambda_g=0.05$, keeping both penalties below typical per-slot key delivery while still exposing handover cost to the learner.

\begin{algorithm}[t]
\caption{\textsc{SATLOCK} Online Routing per Orbital Window}
\label{alg:framework}
\begin{algorithmic}[1]
\Require Orbital data, weather model, trained agents $\{\pi_k\}$
\For{each time slot $t = 1, \ldots, T$}
  \State Update satellite positions and visibility matrix $\mathbf{V}_t$
  \State Sample weather state via Markov chain~(\ref{eq:weather})
  \For{each demand pair $(s_k, d_k) \in \mathcal{K}$}
    \State Compute feasible paths: direct and 1-hop relay
    \State Construct state $\mathbf{s}_t^k = [t/T,\,\mathbf{V}_t^k,\,\mathbf{w}_t^k,\,\mathbf{b}_t^k,\,g_t^k]$
    \State $a_t^k \gets \pi_k(\mathbf{s}_t^k)$ \Comment{Agent selects path type}
    \State Tentatively assign $R_{k,t,a_t^k}$ key bits to demand $k$
  \EndFor
  \State Resolve conflicts: if satellite $s$ is claimed by multiple demands,
         retain the assignment with highest $R_{k,t,a_t^k}$; others skip slot $t$
  \State Accumulate delivered key bits: $\mathbf{b}_{t+1}^k \gets \mathbf{b}_t^k + R_{k,t,a_t^k}$
\EndFor
\end{algorithmic}
\end{algorithm}

\subsection{DQN Architecture}
\begin{figure*}[hbt]
  \centering
  \includegraphics[width=0.75\textwidth]{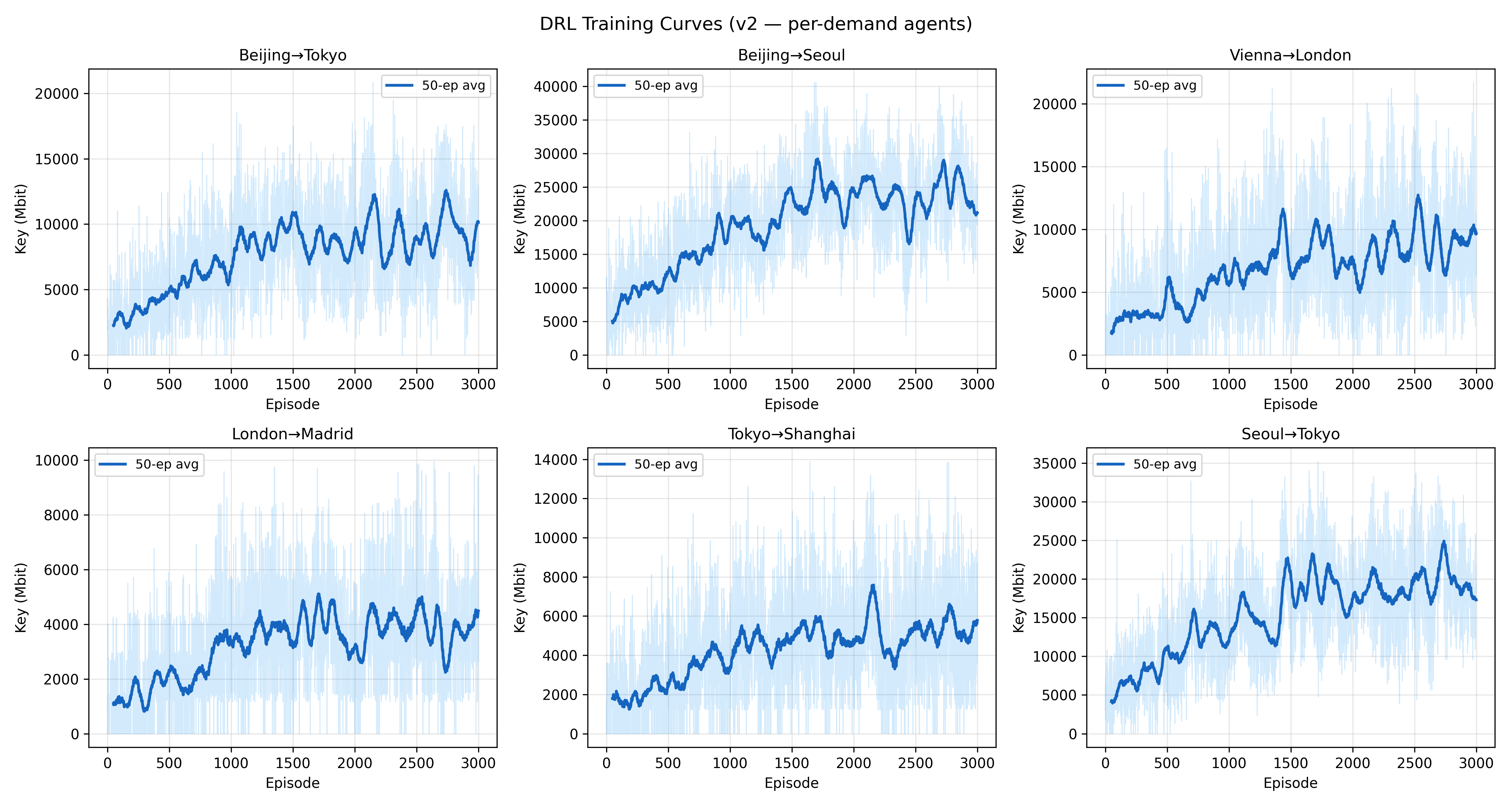}
  \caption{DRL training curves for six demand pairs. All agents converge by episode 3000; the $\varepsilon$-greedy policy produces noisy early
    rewards that smooth into a stable plateau.}
  \Description{DRL training curves showing convergence of all agents by episode 3000.}
  \label{fig:training}
\end{figure*}

Each agent is a 3-layer multi-layer perceptron (MLP; two hidden layers of 128 units, ReLU activations) trained with experience replay (buffer 10\,000, batch~64) and a target network synchronized every 50 episodes, using the Adam optimizer ($\text{lr}=3\times10^{-4}$), discount factor $\gamma=0.95$, and $\varepsilon$-greedy exploration decaying from $1.0$ to $0.05$ over the first 3000 of 5000 training episodes. At inference, each agent selects an action in ${\sim}50\,\mu$s, roughly ${16{,}000\times}$ faster than re-solving the ILP (${\sim}0.8$\,s), enabling real-time slot-by-slot deployment. Figure~\ref{fig:training} shows convergence for all six demands.

\section{Evaluation}
\label{sec:eval}

\subsection{Experimental Setup}
Table~\ref{tab:params} lists all simulation parameters. We compare six algorithms: \textbf{ILP} (optimal upper bound), \textbf{DRL} (trained DQN), \textbf{Greedy} (max-rate path per slot), \textbf{Caleffi}~\cite{caleffi2017optimal} (Werner-fidelity product), \textbf{Pant}~\cite{pant2019routing} (transmittance product), and \textbf{Direct-only} (no relay). The Caleffi and Pant criteria are re-implemented as per-slot path-scoring rules on our channel model, so all six methods consume the identical finite-key rate function and stochastic channel realizations; Greedy, Caleffi, Pant, and Direct-only are myopic per-slot policies, while only the ILP optimizes across slots. Results are means over 30 held-out weather seeds.

\begin{figure}[t]
\centering
\includegraphics[width=\linewidth]{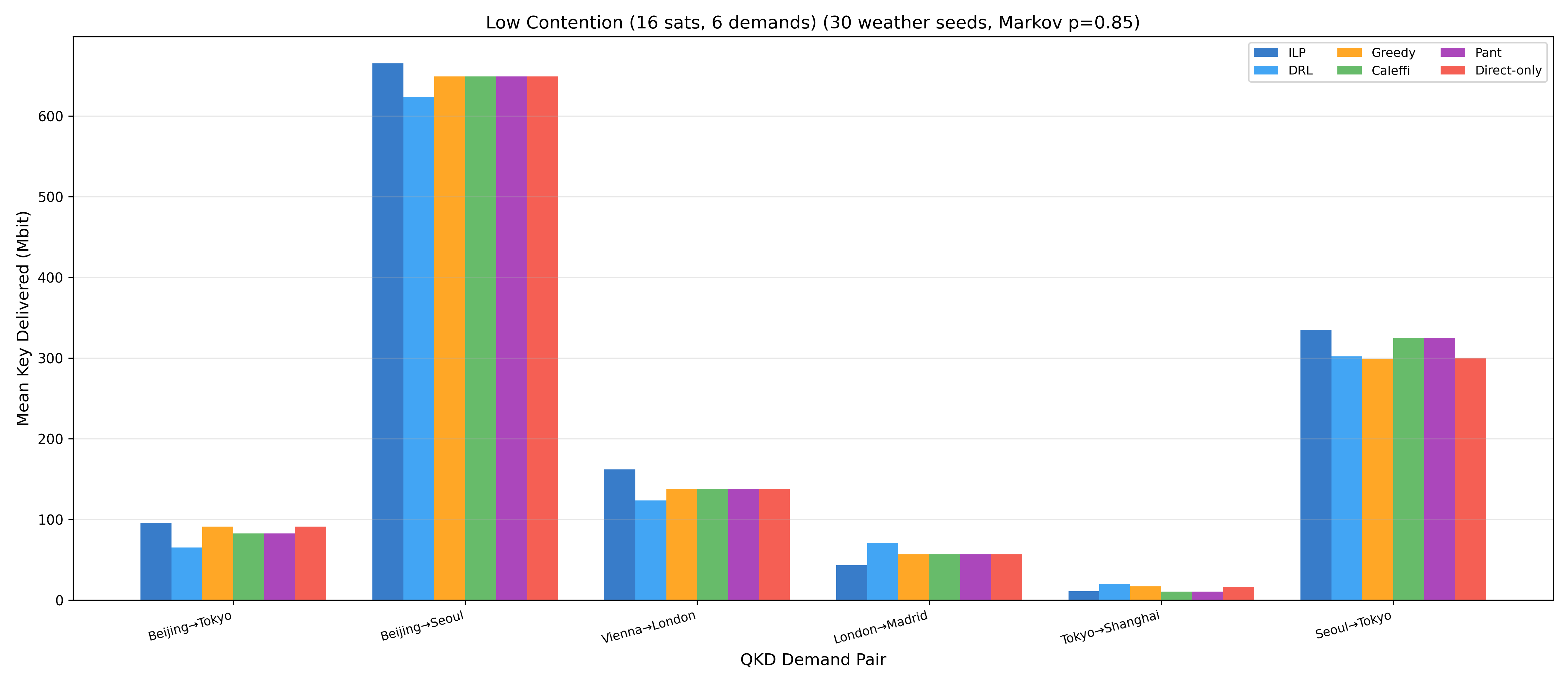}
  \Description{Grouped bar chart of per-demand throughput for the six routing algorithms under low contention (16 satellites, 6 demands).}
\vspace{-3pt}
\caption{Low-contention per-demand throughput (16 satellites, 6 demands, 30 seeds).}
\label{fig:vsgreedy}
\end{figure}

\begin{figure}[b]
\centering
\includegraphics[width=\linewidth]{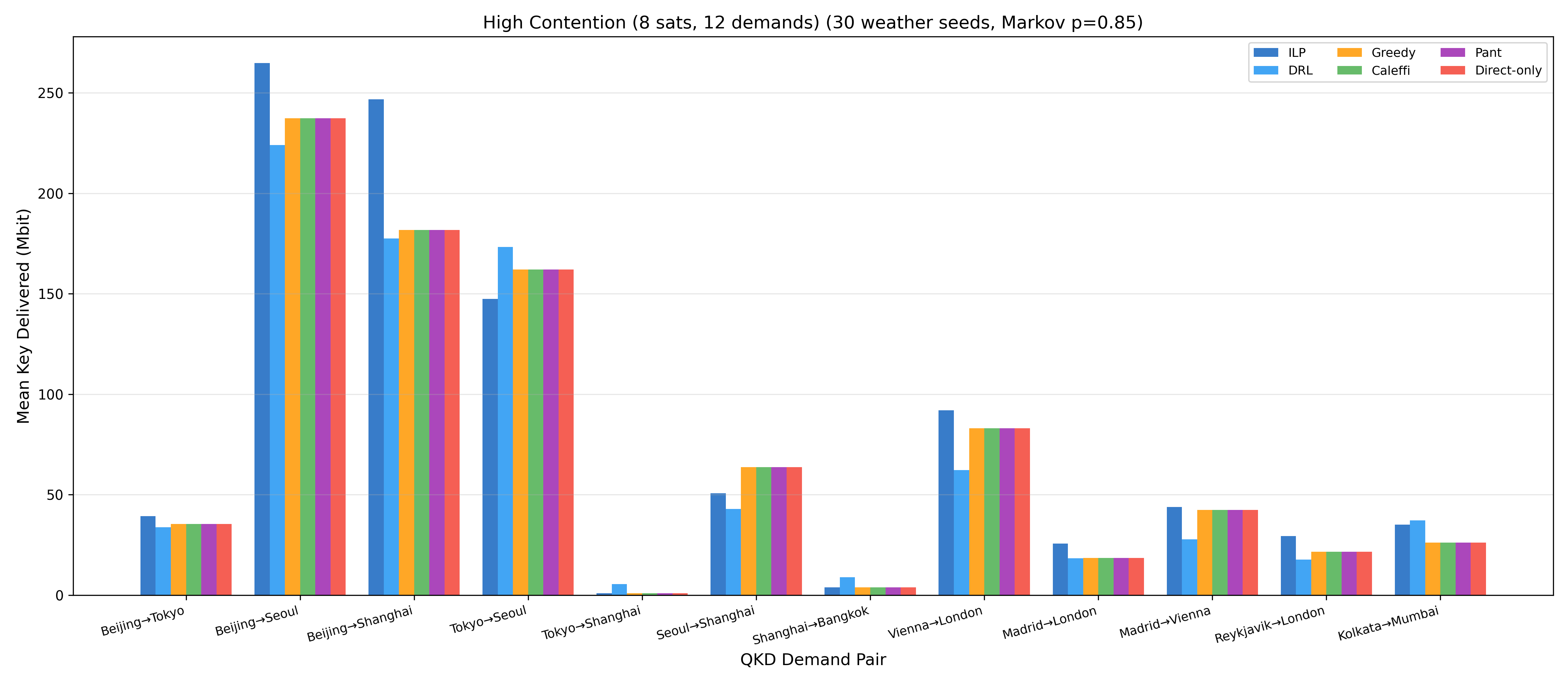}
  \Description{Grouped bar chart of per-demand throughput for the six routing algorithms under high contention (8 satellites, 12 demands).}
\vspace{-3pt}
\caption{High-contention per-demand throughput (8 satellites, 12 demands, 30 seeds).}
\label{fig:highcontention}
\end{figure}

\begin{table}[b]
\caption{Simulation Parameters}
\label{tab:params}
\centering
\footnotesize
\setlength{\tabcolsep}{4pt}
\renewcommand{\arraystretch}{1.1}
\begin{tabular}{@{}llll@{}}
\toprule
\textbf{Parameter} & \textbf{Value} & \textbf{Parameter} & \textbf{Value} \\
\midrule
Ground stations & 10 (Asia+Eur.) & Pulse rate $f_{\rm rep}$ & 100\,MHz \\
LEO satellites & 4--16 & Detector eff.\ $\eta_{\rm det}$ & 0.20~\cite{liao2017satellite} \\
Planes / incl.\ & 4, 53$^\circ$ & Dark count $d_c$ & 300/s (InGaAs) \\
Altitude $h$ & 500\,km & Basis sift $f_{\rm sift}$ & 0.5 (BB84) \\
Min.\ elev.\ $\theta_{\min}$ & 10$^\circ$ & Handover loss $\alpha$ & 0.5 \\
Duration/slot $\Delta t$ & 120/2\,min & DRL episodes/hidden & 5000/128$\times$2 \\
Pointing jitter $\sigma_p$ & 0.7\,$\mu$rad & Replay buf./batch & 10\,000/64 \\
Security $\varepsilon_s$ & $10^{-10}$ & Learning rate & $3\times10^{-4}$ \\
Weather persist.\ $\rho$ & 0.85 & & \\
\bottomrule
\end{tabular}
\end{table}

\subsection{Per-Demand and Algorithm Comparison}
\sloppy
Under Markov-correlated weather and handover loss (30~held-out seeds), the algorithms separate as contention increases. In low contention (16~satellites, 6~demands), the strongest heuristics remain close to ILP since satellite exclusivity is only occasionally binding; even there, the handover-aware ILP delivers the largest aggregate key volume (1,311~Mbit, vs.\ 1,262~Mbit for Caleffi/Pant, 1,250~Mbit for Greedy, and 1,204~Mbit for DRL; Table~\ref{tab:algcomp}). The DRL agent trails the best heuristics in this regime, the price of decentralized per-demand decisions; its value lies in microsecond inference and weather adaptivity rather than low-contention throughput. The per-demand breakdown in Figure~\ref{fig:vsgreedy}, computed under the same Markov weather and handover loss, also exposes a fairness behavior: since~(\ref{eq:obj}) maximizes total delivered key volume, it favors less-impaired demand pairs, visible in the large gap between high-volume pairs like Beijing--Seoul and the near-zero delivery to Tokyo--Shanghai, and can leave the weakest underserved under high contention. We optimize aggregate throughput because delivered bits are the security-relevant resource and total throughput is a single well-defined certifiable target; this trades per-demand fairness, and could in principle starve a persistently cloud-blocked station. A fairness-aware variant, e.g., max-min or proportionally fair objectives, or a per-demand minimum-key constraint, is a natural extension, likely trading some aggregate volume for more even delivery; we leave its evaluation to future work.

\begin{table}[t]
\caption{Algorithm Comparison: Total Key Delivery (Mbit, 30 seeds).
Low: 16 satellites, 6 demands. High: 8 satellites, 12 demands.
\% computed as algorithm total / ILP Optimal $\times$ 100.}
\label{tab:algcomp}
\centering\small
\setlength{\tabcolsep}{4pt}
\begin{tabular}{lcc|cc}
\toprule
& \multicolumn{2}{c|}{\textbf{Low Contention}} & \multicolumn{2}{c}{\textbf{High Contention}} \\
\textbf{Algorithm} & \textbf{Key} & \textbf{\%} & \textbf{Key} & \textbf{\%} \\
\midrule
ILP Optimal                       & 1,311 & 100.0\% & 980 & 100.0\% \\
Caleffi~\cite{caleffi2017optimal} & 1,262 & 96.3\%  & 877 & 89.5\%  \\
Pant~\cite{pant2019routing}       & 1,262 & 96.3\%  & 877 & 89.5\%  \\
Direct-only                       & 1,251 & 95.4\%  & 877 & 89.5\%  \\
Greedy (max-rate)                 & 1,250 & 95.3\%  & 877 & 89.5\%  \\
DRL Agent                         & 1,204 & 91.8\%  & 829 & 84.6\%  \\
\bottomrule
\end{tabular}
\end{table}

The separation is larger under high contention (8~satellites, 12~demands), where satellite conflicts and handovers occur more frequently: as Figure~\ref{fig:highcontention} shows under the same 30-seed evaluation, the ILP delivers 980~Mbit, the myopic baselines drop to 877~Mbit (89.5\% of ILP), and the decentralized DRL agent delivers 829~Mbit (84.6\%; Table~\ref{tab:algcomp}). Here, constraint C2 binds at nearly every slot, forcing all myopic algorithms to the same feasible assignment and identical throughput, visible as matching bar heights across baselines for most demand pairs in the figure. This confirms the paper's central claim: once handover overhead creates temporal coupling, globally optimized scheduling measurably beats per-slot myopic routing. The DRL result is more nuanced: independent agents learn useful weather-sensitive policies on some pairs, but aggregate performance is limited by the lack of explicit multi-agent coordination under shared satellite constraints.

\subsection{Global Routing Under Weather Uncertainty}
\begin{figure}[t]
  \centering
  \includegraphics[width=\columnwidth]{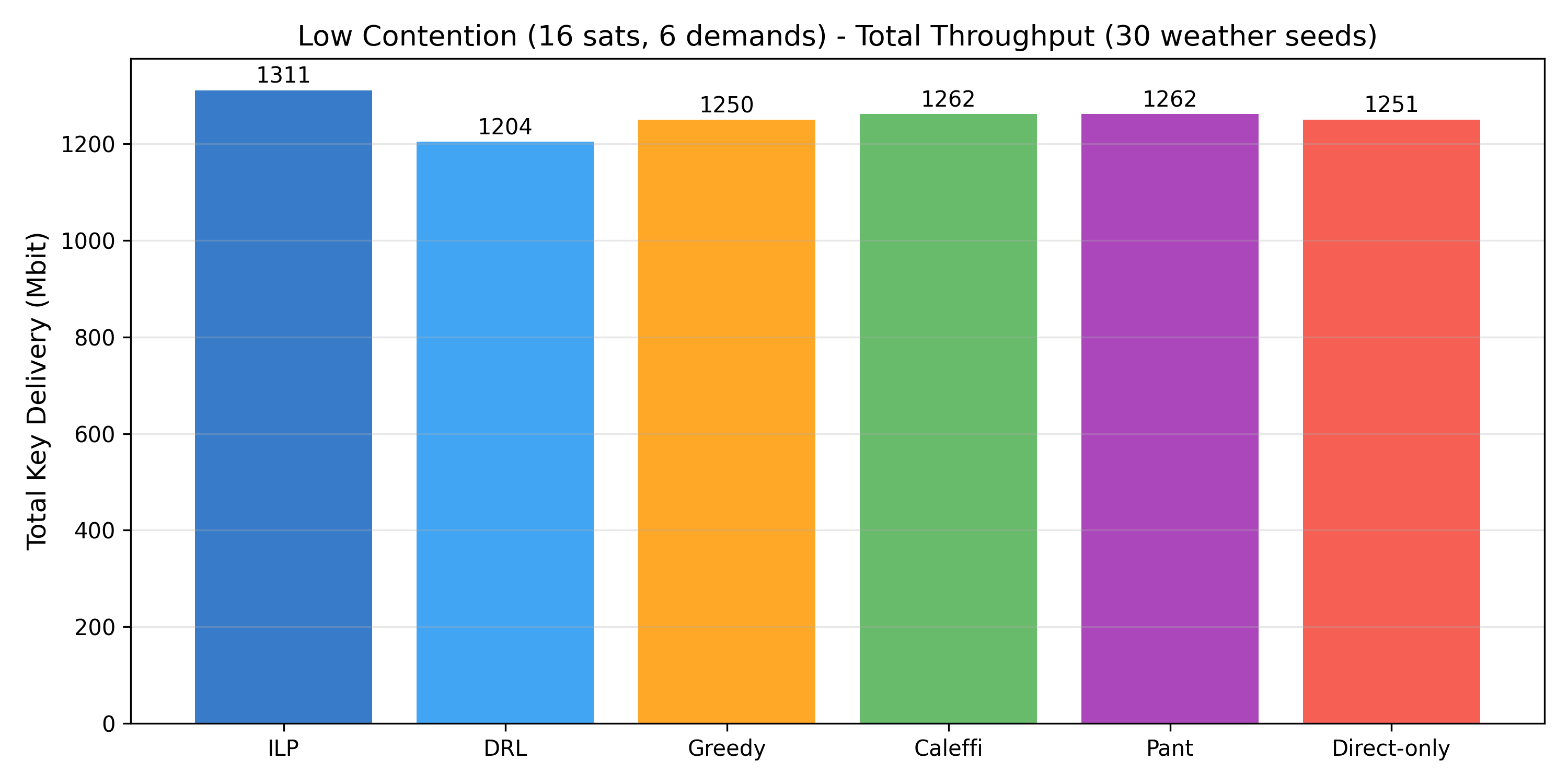}
  \Description{Bar chart of total delivered key bits per algorithm under stochastic weather across contention regimes.}
  \caption{Aggregate key delivery, low-contention constellation (30 seeds).}
  \label{fig:global}
\end{figure}

Figure~\ref{fig:global} shows aggregate key delivery across all six routing algorithms on the 16-satellite constellation with stochastic Markov weather and handover loss (30~seeds). Averaging over weather realizations isolates the algorithmic contribution from channel luck: the ordering of methods is stable across seeds, indicating that the ILP advantage stems from handover-aware scheduling rather than favorable weather draws. The handover-aware ILP benefits from planning across consecutive slots instead of greedily reassigning satellites whenever the instantaneous rate changes. The decentralized DRL agent discovers demand-specific allocation strategies, but its independent action selection cannot fully coordinate satellite conflicts across demands. In deployment terms, the two ends of this spectrum are complementary: the ILP suits offline pass-plan generation at the network operations center, while the DRL agent suits onboard or edge execution when weather deviates from the plan and re-solving is impossible within a slot.

\subsection{Discussion}
The handover-aware ILP solves in ${\sim}0.8$\,s at this scale (64+ satellites would need column-generation or Lagrangian decomposition~\cite{pirandola2020advances}), and handover loss couples adjacent decisions, exposing the value of global scheduling over per-slot selection. The DRL agent infers in ${\sim}50\,\mu$s per slot without re-solving an NP-hard program, but its independent-agent design cannot fully resolve centralized satellite conflicts~\cite{MAH_decentralized}. Per-demand agents avoid the combinatorial joint action space ($3^K$) and specialize to each pair's conditions at this coordination cost.

Two design choices merit emphasis. Precomputing the best path per type and choosing only among types reduces the action space to 3 discrete actions, enabling convergence within 3,000 episodes. Meanwhile, including the cumulative key progress $b_t$ and a handover flag $g_t$ in the state enables demand-aware allocation, redirecting capacity to underserved pairs once others are satisfied. Both insights may transfer to other quantum network problems with combinatorial path spaces and stochastic channels. The Markov weather model ($\rho=0.85$) matches typical ECMWF persistence statistics and can be replaced with real per-station climatologies directly, since the optimization and agent consume only the resulting per-slot rates; finite-key effects (Eq.~\ref{eq:fk}) eliminate sub-60\,s passes entirely, confirming the value of physically faithful channel models~\cite{ribezzo2023deploying}.

Our evaluation remains simulation-based, with cloud cover drawn from a two-state Markov process using representative climatological parameters rather than measured traces, which is the study's main limitation. Validating against real meteorological traces and recorded pass geometries is our immediate next step, with centralized or communication-enabled multi-agent DRL as the natural path for closing the gap to ILP.

\section{Conclusion}
\label{sec:conclusion}
We presented \textsc{SATLOCK}, a handover-coupled scheduling framework for LEO QKD networks that combines a physically grounded channel model (finite-key correction, decoy-state estimation, Markov weather, pointing jitter, and handover loss) with a provable ILP optimality certificate and a weather-adaptive online routing agent. The evaluation shows contention-dependent algorithm differentiation: the best heuristics reach 95--96\% of ILP under low contention but drop to 89.5\% under high contention as satellite conflicts and handover penalties become binding, while the decentralized DRL agent learns useful weather-sensitive demand policies but is limited in aggregate by independent per-demand coordination. Future work includes centralized multi-agent DRL, multi-hop entanglement routing with quantum repeaters, per-station integration of real ECMWF cloud-cover data, and hardware-in-the-loop validation using SDR testbeds or emulated QKD devices, extending toward end-to-end network architectures with AI-assisted slicing and resource management~\cite{MAH_ISAC}.

\bibliographystyle{ACM-Reference-Format}
\bibliography{reference}

\end{document}